\documentclass[aps,preprint]{revtex4}
\usepackage{graphicx}
\usepackage{color}
\begin{document}

\title{Giant dielectric permittivity and magneto-capacitance effects
in low doped manganites}
\author{R.F Mamin$^{1,2}$, V.V. Kabanov$^{2}$}
\affiliation{$^{1}$Zavoisky Physical-Technical Institute, Kazan Scientific Center of Russian Academy of Sciences, 420029 Kazan, Russia}
\affiliation{$^{2}$Complex Matter Department, Jozef Stefan Institute, SI-1000, Ljubljana, Slovenia}
date{\today}

\begin{abstract}
The effect of giant dielectric permittivity due to phase separation accompanied by the charged inhomogeneities in the low doped manganites is discussed. The effect appears in the vicinity of the second order magnetic phase transition which is caused by the long range Coulomb forces. The long range Coulomb interaction is responsible for the formation of the inhomogeneous charged states and determines the characteristic length scales. We derive the phase diagram of the inhomogeneous charged states in the framework of the phenomenological theory of phase transitions. The large value of static dielectric function reduces characteristic value of the Coulomb energy of the inhomogeneous state and makes the appearance of the magnetoelectric effect possible. We discuss the formation of state with giant dielectric permittivity and magneto-capacitance effects in that case.
\end{abstract}


\maketitle

The importance of the inhomogeneous phase segregated states
in the explanation of the anomalous transport and magnetic properties
in
manganites \cite{jin,tokura,millis,nagaev,dagotto,dagotto1,ramirez,coey, kagan1,kagan2}
and in
high temperature superconductors (HTS)
\cite{gorkov,emery1,grisha} is widely discussed.
In doped manganites the complex interactions between different
degrees of freedom lead to unusual magnetic and transport
properties. Moreover it was suggested\cite{most} that spiral
magnetic order observed in un-doped manganites $\rm ReMnO_3$ may
lead to multiferroic behavior. It was also
discussed
\cite{Khoms}
that the charge ordering in magnetic systems may cause the
magnetoelectric effect. Here we propose that the magnetoelectric
effect may appear not due to charge ordering but due to charge
segregation, which appears near Coulomb frustrated second order
phase transition.
In this paper we discuss the tendency and conditions of the
formation of inhomogeneous states with the spacial charge
localization and the phase separation within the phenomenological
theory and clarify the role of the Coulomb interaction in this
phenomenon. We demonstrate the possibility of
the magnetoelectric
behavior in the low doped
manganites
and clarify the role of the Janh-Teller interaction in that behavior.

The problem of Coulomb-frustrated phase separation in different
charged systems is the subject of ongoing discussion
\cite{millis,nagaev,dagotto,dagotto1,gorkov,emery1,grisha,emery3,spivak,jamei,di_castro_1,Lorenz_2,mertelj,fine}.
Numerous studies have focused on first-order phase transitions
where the charge density is coupled linearly to the order
parameter (as an external field)~\cite{jamei,di_castro_1} or to
the square of the order parameter (local
temperature)~\cite{mertelj}.

The importance of the Coulomb interaction in the formation of
 inhomogeneous charged states in the doped manganites was previously emphasized
\cite{Shenoy_1,Shenoy_2, Shenoy_3,Kugel_jetp,Kugel_prb2006,Kugel_prb2007}.
In many cases
\cite{Shenoy_1,Shenoy_2,Shenoy_3,Kugel_jetp,Kugel_prb2006,Kugel_prb2007}
in order to describe the phase separated state near the magnetic phase transition an interaction with additional degrees of freedom were considered.
The interaction with these degrees of freedom leads to the energy gain compared to the case of the purely magnetic phase separation. This energy gain is relatively small (about $3kT_c\simeq$  0.03-0.1 eV per one unit cell). Therefore this type of phase separation is not plausible.
It follows from the fact that characteristic energy of magnetic interaction $kT_c\simeq$  0.01-0.03 eV is less then the Coulomb energy $V_c\simeq$ 0.10-0.13 eV. Here we use the static dielectric constant for manganites $\varepsilon = 30-40$.
In order to justify the existence of the phase segregation near the magnetic phase transition
in the presence of the long-range Coulomb interaction
some authors\cite{Shenoy_1,Shenoy_2,Shenoy_3,Kugel_jetp,Kugel_prb2006,Kugel_prb2007} underline the important role of the Jahn-Teller effect, which is characterized by the Jahn-Teller energy $J_{JT}$. According to
Refs.\cite{Shenoy_1,Shenoy_2,Shenoy_3,Kugel_jetp,Kugel_prb2006,Kugel_prb2007}
the main contribution to the energy of the low-temperature phase
is due to the Jahn-Teller distortions.
For example in Refs.\cite{Shenoy_1,Shenoy_2} the nanoscale electronic inhomogeneities
are discussed for $J_{JT}\simeq$ 0.5 eV. In our view in Refs.\cite{Shenoy_1,Shenoy_2,Shenoy_3} the characteristic energy of the Coulomb
energy is underestimated $V_c \simeq$0.02 eV (for $\varepsilon = 20$) at least by the order of magnitude.
But what is more important that in manganites the phase separation in the vicinity of the
magnetic phase transition is observed. Therefore in order to make correct conclusions about phase separated state we should compare the contribution of the relevant interactions to the total
energy of the inhomogeneous state rather then different coupling constants.
In this paper we demonstrate that the phase separated state accompanied by the charged inhomogeneities in manganites arises naturally near the magnetic phase transition frustrated by the long range Coulomb forces without coupling with any additional degrees of freedom like Jahn-Teller distortions.
We underline that for the existence of this effect it is important to have relatively strong interaction between the electrons and magnetic degrees of freedom. Another important  conclusion is that the phase separation becomes plausible due to the self-screening of the space-charge inhomogeneities provided that the screening due to the lattice charges is strong leading to the large values of the dielectric permittivity $\varepsilon $= 30-40.
We show that in the absence of the percolation between phase
segregated regions there exists a large contribution to the dielectric constant which is
connected with the displacements from equilibrium positions of the charged clusters. This
contribution to the dielectric function is magnetic field dependent and leads to the
magnetoelectric effect. Therefore we expect the giant dielectric permittivity
and magnetocapacitance effects in low doped manganites.

We accurately analyze the total energy of the system and
treat Coulomb interaction exactly because it is the key factor
which determines the characteristics of inhomogeneous states. The
idea of phase segregation and inhomogeneous charge distribution in
manganites was successfully applied for the description of the
magnetoresistive effect in the limit of the percolating charged
regions \cite{tokura,millis,nagaev,dagotto,dagotto1}. In this paper
we emphasize the possibility of magnetocapacitance effect due to
polarization of the non-percolated nano-regions in external
electric field in the inhomogeneous state in the low-doped
systems.

The phenomenological approach to the theory of Coulomb frustrated
phase transition
emphasize
that
the
properties
the system are universal and
are determined
by the closeness to the phase transition point and
by the
dependence of
the critical temperature of the phase transition
on doping. This approach is essentially
independent on other properties of the system.
Therefore,
the results
demonstrate
that the properties of the
inhomogeneous states
are determined by the proximity to the phase
transition and the strength
of the Coulomb interaction interactions
but
are independent on
other microscopic interactions in manganites.

Our approach indicates that the phase separation with the formation of charged inhomogeneities  is a quite common phenomenon inherent to a various systems with the different types of the phase transitions, such as manganites and HTS materials.
Our estimates show that the Coulomb energy
in  the charged separated states is
relatively small.

We consider a doped system in the vicinity of a second order
magnetic phase transition. We assume that the free
carrier density $\rho$ is proportional to the dopant concentration
$x=\rho /\rho_0$, where $\rho_0=e/a^3$ characteristic carrier
density,
$e$ is elementary charge, and $a$ is the lattice parameter.
The thermodynamic potential $\Phi=\int\phi(\eta,\rho)d^3r$
describes the behavior of the magnetic order parameter $\eta$ near the
second
order phase transition and the coupling between the order
parameter
and the charge density reads:
\begin{eqnarray}
&&
\phi(\eta,\rho) = \phi_{0}+\phi_{\eta}+\phi_{int}+\phi_{e}
\\
&&
\phi_{\eta}(\eta) = \frac{\alpha}{2}\eta^{2}+
\frac{\beta}{4}\eta^{4}+ \frac{\xi}{2}(\nabla\eta)^{2}-\eta H
\nonumber\\
&&
\phi_{int}(\eta,\rho)=-\frac{\sigma}{2}\eta^{2}
\frac{\rho(\rho_0-{\rho})}{\rho_0^2}
\nonumber
\\
&&
{\phi _{e}}(\rho ) =
\frac{\gamma({\rho (r)}-{\bar{\rho}})}
{2\rho_0^2}
\int\frac{({\rho (r')}-{\bar{\rho }})}{|r-r'|} d V'
+
\frac{\vartheta({\rho (r)}-{\bar{\rho}})^2}{2\rho_0^2}
\nonumber
\end{eqnarray}
where $\phi_{0}$ is the density of the thermodynamic potential in
the high-temperature phase, $\phi_{\eta}$ is the density of the
thermodynamic potential of the low-temperature phase, $\alpha$,
$\beta$, $\xi$ are coefficients in the expansion of the
thermodynamic potentials in powers of the order parameter
($\alpha=\alpha^{'}(T-T_{c})$, where $T_{c}$ is critical
temperature
in the absence of doping, $\alpha^{'}=1/C$, $C$ is the Curie
constant), $\xi$ is proportional to the co-called diffusion
coefficient of the magnetization $\eta$ and is defined by the exchage ineraction.
It defines the characteristic length $\xi^{1/2}$ where the order parameter changes.
$H$ is the external magnetic field.
$\phi_{int}$ describes the interaction of the order
parameter with the charge density, $\sigma$
is the constant of interactions. In order to provide
global stability of the system we require $\sigma>0$
\cite{emery3}.
$\phi_{e}$ describes the charging effects due to Coulomb
interaction between
the charge carriers, and $\bar{\rho}$ is the average charge
density.
The spatial distribution of the charge density $\rho$ and the order 
parameter $\eta$ are determined from minimization of the thermodynamic 
potential. Therefore the effects of screening are calculated self-consistently 
during minimization of the thermodynamic potential (1).
The inclusion of the gradient term for the charge density is not necessary, because the spatial disribution of the charge density is determined by the minimization of the Coulomb energy.
We assume that Coulomb contribution is the strongest:
$\vartheta \ll a^2 \gamma $, where $\vartheta$ is the constant that describes the strength
of the local electron-electron interaction, differnt from the Coulomb repulsion.
The effect of average dopant concentration $\bar{\rho}$ to the
thermodynamic potential in the high-temperature phase is included
in $\phi_{0}$. The coefficient $\gamma$ is inversely proportional
to the static dielectric constant $\varepsilon$,
$\gamma=\rho_0^2/\varepsilon$.
The expression for $\gamma$ takes into account the effect of the local excess 
positive charge of the lattice explicitly via static dielectric constant. 
Indeed the direct Coulomb repulsion between electrons in crystals is reduced 
due to electronic polarizability by the factor $\epsilon_{\infty}$. If we take 
into account polaronic effects effective interaction between electrons is 
renormalized and $\epsilon_{\infty}$ should be replaced by the static dielectric 
function (for details see the Ref. \cite{AlexandrovKabanov}).
As a result the coefficients in the thermodynamic potential depend
on
charge density $\rho$ and the term
$\sigma\rho(\rho_0-\rho)$ determines
the shift of the critical temperature $T_{c}$ due to
variation of the local charge density:
\begin{equation}
       T_{c\rho}(x)=T_{c0}+\frac{\sigma x(1-x)}{\alpha^{'}}
\end{equation}
Here  $x=\rho /\rho_0$ and $\bar{x}=\bar{\rho} /\rho_0$
are dimensionless carrier densities.
In the uniform case we have $x=\bar{x}\ (\rho=\bar{\rho})$ and
$\eta_{u}=H/2 \tilde{\alpha}(\bar{x},T)$ for $T>T_{c0}$ and
$\eta_{u}=-\tilde{\alpha}(\bar{x},T)/\beta-
H/\tilde{\alpha}(\bar{x},T)$
for $T<T_{c0}$ ($\tilde{\alpha}(x,T)=\alpha(T)-
{\sigma}x(1-x)$).

Note that we consider the inhomogeneous states, which appears near the
second order phase transition to the ferromagnetic phase. Therefore we do not consider any effect
 of antiferromagnetic phase or charge ordering. We believe that the influence of the
 antiferromagnetic or the charge ordered states to the formation of the charge segregated state is
 not important.
We assume that magneto-dipole coupling between bubbles is small in comparison with the exchange interaction. The tunneling of the carriers is important because it leads to the exchange interaction between bubbles. This leads to the magnetic ordering of different bubbles, divergence of the correlation radius and appearance of the macroscopic magnetization in the sample. Therefore we consider the temperature $T_{c\rho}(x_1)$ (2) as the temperature of the phase transition in the system of bubbles. 

In Fig.1 we plot the free energy as a function of the carrier density $x$ for case of the uniform
 phases. In the region of density where the second derivative of the free energy on $x$ is
 negative,
the uniform state becomes metastable. The phase segregated nonuniform states have lower energy as
demonstrated in Fig.1. Therefore there is a tendency to phase separation into domains
with different carrier density $x$ (for details see Ref.\cite{fine}).
%
%
\begin{figure}[b]
\begin{minipage}{1\linewidth}
\includegraphics[width=0.7\linewidth]{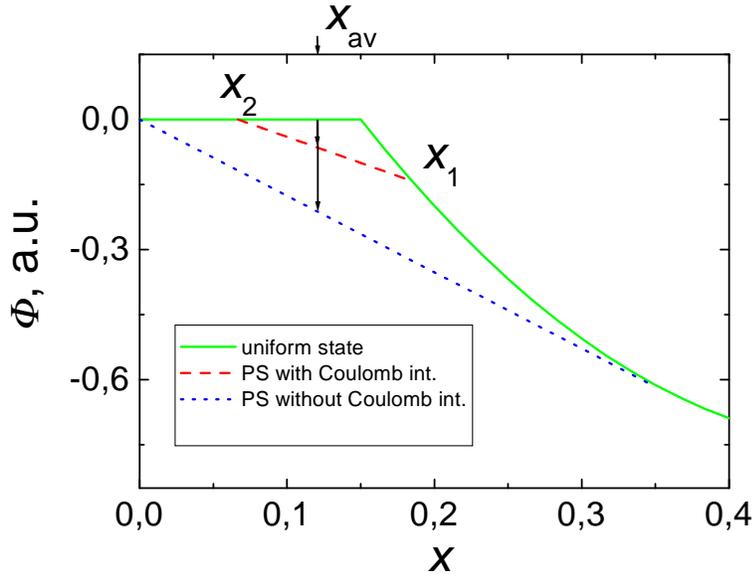}
\caption{The dependence of the total energy of the uniform state (solid line) on the carrier density $x$.
The arrow on the dotted line shows the energy of the phase separated state for case
without Coulomb
interaction. The arrow on the dashed line indicate the energy of the phase separated state
for case where the contribution of the Coulomb interaction is considered. The energy is calculated for the case $\bar{x}=x_{av}$, $T_{c0}=-60K$,
$\sigma/\alpha^{'}=1750K$ for $T= constant$. The parameters $x_{1}$ and $x_{2}$
correspond to the carrier concentrations in the ferromagnetic and in the paramagnetic phases.}
\label{fig:0}
\end{minipage}
\end{figure}
In the equilibrium state the order parameter is determined
as a steady state solution of the Landau-Khalatnikov
equation $\partial\eta/\partial t=\Gamma \delta \phi(\eta,x)/\delta\eta$
\cite{Khalatnikov}: $\delta \phi(\eta,x)/\delta \eta=0$. The carrier
density is determined from the following equation:
$\delta \phi(\eta,x)/\delta x=\mu_e$,
$\mu_e$ is the
chemical potential,  which can be found
from the conservation of the total number of carriers.
%
%
The
second equation reads:
\begin{eqnarray}
-\frac{\sigma}{2}\eta^{2}
(1-2x(r))
+\gamma
\int\frac{({x(r')}-{\bar{x}})}{|r-r'|} d V'
+\vartheta ({x(r)}-{\bar{x}})=0
%
%
\end{eqnarray}
\noindent
Using the formula $4\pi \nabla^2 |r-r'|   = - \delta (|r-r'|)$  and
assuming that $\vartheta \ll a^2 \gamma $,
we obtain the dependence of the equilibrium charge density $x_{s}$ on
$\eta$:
\begin{eqnarray}
x_s (\eta) \simeq \bar{x} & -& \frac{\sigma (1 - 2 \bar{x} )}{8
\pi\gamma} \Big( {\nabla}^2 \eta^2 - \frac{\sigma}{4 \pi\gamma}
{\nabla}^2[\eta^2 {\nabla}^2 \eta^2]
\Big)\\
&-& \frac{\vartheta \sigma }{64 \pi^2\gamma^2} \nabla^4 \eta^2
\nonumber
\end{eqnarray}

\noindent
Substituting Eq.(4) to Eq.(1)
$(\vartheta \ll a^2 \gamma )$
we obtain the expression of the density of the thermodynamic potential (1)
$\phi_{\rho s}$ for the equilibrium distribution of the
charge density $x_{s}(\eta)$:
\begin{eqnarray}
\phi_{\rho s}(\eta)  =       \phi_{\eta}(\eta) -
\frac{\sigma\bar{x}(1-\bar{x})}{2}\eta^{2} -
\frac{\sigma^2(1-2\bar{x})^{2}} {32\pi \gamma} (\nabla
\eta^{2})^{2}&&
\\
+
\frac{\vartheta \sigma^2(1-2\bar{x})^{2}}{256 \pi^2 \gamma^2}
(\nabla^2 \eta^2)^2 + \frac{\sigma^3(1-2\bar{x})^{2}} {128\pi^2
\gamma^2}  \eta^{2}(\nabla^{2}\eta^{2})^{2}.&& \nonumber
\end{eqnarray}
The negative sign in the third term of Eq.(5)
indicates that the uniform state may
be unstable towards inhomogeneous fluctuations.

This instability leads to spatially inhomogeneous solutions.
Calculation of the total thermodynamic potential of the inhomogeneous state including the Coulomb
energy and the energy interphase boundaries shows that the minimum of the thermodynamic potential
corresponds to the phase separated state.
Local minima may correspond to different inhomogeneous states where new phase is organized in the form of symmetric bubbles or the periodic stripes or some other arrangements.
The
most
simple solution is spherically symmetric bubbles of the low
symmetry
phase separated by the large distance from each other.

This type of solution (charged bubble of magnetic phase screened
by external charge)
represent the minimum of the Coulomb energy
together with the energy of the interphase boundaries.
Therefore this phase separated state has the lowest energy through the most part of the phase diagram and in particular
in the vicinity of
the upper boundary of the appearance of the inhomogeneous states\cite{Lorenz_2}.
Thus this equilibrium inhomogeneous state in the low doped samples
has
the spherical form and the distribution of charge has a shape of
electric double-layer \cite{KM_jetf}.
This solution has
characteristic size $R_{0}$. The characteristic
average charge density inside of bubbles is $x_{1}$
and the average charge density outside of these regions is $x_2$.
The average charge density in the system is $\bar{x}$.
The average value of the order parameter inside of the bubble
$\eta_1\simeq \eta_{0}
+H/2(-\tilde{\alpha}({x_1},T))$ is relatively large, while
the order parameter outside of bubbles is much smaller
$\eta_{2}
=H/ \tilde{\alpha}(x_2,T)$
where
$\eta_{0}^2 = -\tilde{\alpha}({x_1},T)/\beta$. It is
clear that the charge is concentrated near the surface of the
sphere. Therefore we can apply the approximation  of the double
electrical
layer for evaluation of the Coulomb energy. At large distances
from the sphere the order parameter and the charge density are
equal to their equilibrium values $x=\bar{x}$ and
$\eta=\eta_{s}(\bar{x})$ ($\eta_{s}\simeq\eta_{2}(\bar{x})
=H/ \tilde{\alpha}(\bar{x},T)$).
Therefore the thermodynamic potential $\Phi_{s}$ of the volume
$V_0$ in that
case has the form:
\begin{equation}
\Phi_{s}
(R_{0},x_{1})
=\Phi_{s0}
-{A(x_{1})\over{3}}R_{0}^{3}+
{B(x_{1})\over{2}}R_{0}^{2}+ {C(x_{1})\over{4}}R_{0}^{4},
\end{equation}
here
$
\Phi_{s0}
=
\phi_{0}V_0-{H^2 V_0
({2\tilde{\alpha}(x_{2},T)})^{-1}},
\
A(x_{1})
=
{\pi\tilde{\alpha}^2(x_{1},T)
\beta^{-1}}
-
{2\pi H^2
\tilde{\alpha}^{-1}(x_{2},T)}
+O(H^3),
\
B(x_{1})
\simeq
8\pi \xi
d^{-1}\beta^{-1}
(-\tilde{\alpha}(x_{1},T)
+ H
\eta_{0}^{-1}(x_{1},T)
+ H^2
(4 \beta \eta_{0}^4(x_{1},T))^{-1}
),
\
C(x_{1})
\simeq
\gamma d
(\bar{x}-x_{1})^{2}F.
$
Here we define dimensionless
factor $F$ ($F=4\int\phi_{e}d^3r/\gamma
R_{0}^{4}d(\bar{x}-x_{1})^{2}$) in order to parameterize the
distribution
of charge. $F=1/18$ in the limit of the
double electrical layer, $V_{0}$ is the volume per
one bubble ($V_{0}=V/n$ where $n$ is the number of
spheres), $d$ is the interphase boundary thickness.
Here we take into account the strong screening of the localized charges
and write the
Coulomb
energy in the double electrical layer approximation.
Internal charge of the double electrical layer represents the charge 
near the surface of the charged bubble. The external charge is of the 
opposite sign and represents the charge of the region where the density 
of the charge carriers is reduced. This charge screens the electric 
field of the charged bubble. As a result the electric field is localized 
in the vicinity of the charged bubble. This distribution of the electric 
field directly follows from numerical minimization of the 
thermodynamic potential (1).
In that case it is proportional to $dR_{0}^{4}$. If the screening
is absent this energy is proportional to $R_{0}^{5}$
\cite{gorkov}.
The parameters $A(x_{1})$ and $B(x_{1})$ (6) depend on temperature
$T$ and magnetic field $H$, and parameter $C(x_{1})$ depends
on the average charge density $\bar{x}$.
Parameters $(\bar{x},T)$ are external and we
find the phase diagram as a function of these parameters. Since
$\bar{x}$ is determined by the doping, we describe the
evolution of the properties of the system with doping.

\begin{figure}[t]
\begin{minipage}{1\linewidth}
\includegraphics[width=0.7\linewidth]{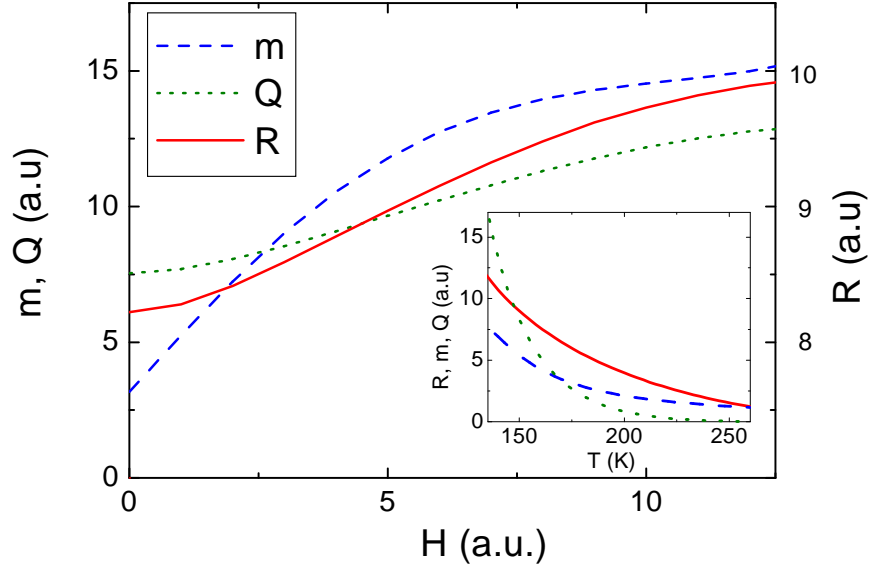}
\caption{Magnetic  field dependence
of $R_s,\ m$ and $Q$. Inset shows the magnetic field dependence
of $R_s,\ m$ and $Q$.}
\label{fig:1}
\end{minipage}
\end{figure}

\begin{figure}[t]
\begin{minipage}{1\linewidth}
\includegraphics[width=0.7\linewidth]{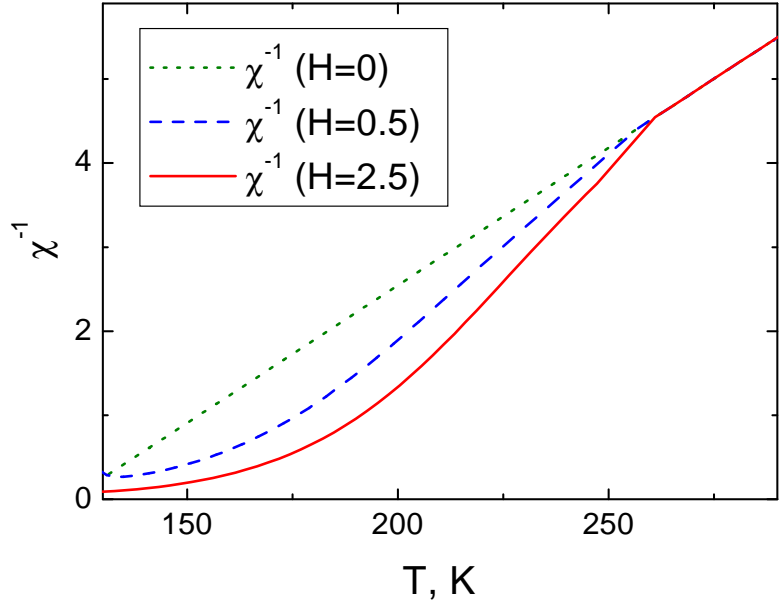}
\caption{The temperature dependence of
$\chi^{-1}(T)$ for two different magnetic fields. }
\label{fig:2}
\end{minipage}
\end{figure}

The minimum of the potential
$\Phi_{s}(R_{0},x_{1})$  (6)
is determined by the set of equations:
$\partial\Phi_{s}(R_{0},x_{1})/\partial R_{0}=0$ and
$\partial\Phi_{s}(R_{0},x_{1})/\partial x_{1}=0$,
which
define
the equilibrium values of parameters: $R_{s}$ and $x_{1s}$.
For the equilibrium size of the
charged domain $R_{s}$ we obtain:
\begin{equation}
R_{s}=\displaystyle\frac{
y_{\eta}
{A_0}(x_{1s})
- {A}(x_{1s})
}
{
(
{y(x_{1s})}^{-1}y_{\eta}
-1)
C(x_{1s})
}
\end{equation}
And $x_{1s}$ is determined by the equation:
\begin{equation}
{\frac{
(y_{\eta}
{A_0}(x_{1s})
- {A}(x_{1s})
)(
(y(x_{1s})
{A_0}(x_{1s})
- {A}(x_{1s})
}
{
(
{y(x_{1s})}^{-1}y_{\eta}
-1)
(y(x_{1s})y_{\eta}^{-1}
-1)
B(x_{1s}) C(x_{1s})
}
}
=1
\end{equation}
\noindent
where we use the notations:
$\eta_3
=
\eta_{0} -
{H}{(-\tilde{\alpha}({x_1},T))^{-1}},\
{A_0}(x_{1})={\pi\tilde{\alpha}^2(x_{1},T)
\beta^{-1}},
\
y(x_{1})=
{4
\sigma(1-2x_1)
(x_1-\bar{x})}
{(-3\tilde{\alpha}(x_{1},T))^{-1}},\
y_{\eta}
=
{4\eta_{1}}
{(3\eta_{3})^{-1}}
$
For computer simulation we assume that $x_{2s}\simeq \bar{x}-
\lambda (x_{1s}-
\bar{x})$. Here $\lambda$
is ratio of the volume of the bubble ($V_1$) and the effective
volume surrounding the bubble ($V_2$),
where the charge density $x$ differs from  $\bar{x}$.
In order to estimate the characteristic length of the
phase segregated regions as well as the screening radius it is necessary to
perform numerical simulations. Nevertheless some estimates may be performed from Eqs.(7,8).
Indeed from Eq.(7) we estimate the typical size of the nano-regions:
$R_{0}\simeq 1-5\ nm$.
Therefore the distance between bubbles may be estimated as:
$L_{R0}\simeq 2(\lambda^{-1/3}-1)R_0$
(here $\lambda  \sim 0.5$ corresponds to the condition
of the compact packing).
The thickness of the interphase boundary is equal to the
the screening length $d$
%
%
should be about $L_{R0}/2$  under the condition
of the compact packing.

As a result for the charge of unite bubble $Q$ we have:
\begin{equation}
Q(T,H) = \rho_0 (x_1-\bar{x}) {4\pi \over 3}R_{s}^3
\end{equation}
Average magnetization $m$ is determined by equation:
\begin{equation}
m(T,H)=\eta_1 \frac{4\pi}{3}\frac{R_{s}^3}{V_0} +\eta_2 \Big( 1-
\frac{4\pi}{3}\frac{R_{s}^3}{V_0} \Big)
\end{equation}
The temperature and magnetic field dependence of $R(T,H),\ Q(T,H)$
and $m(T,H)$
are shown in Fig.2(a,b).
Magnetic susceptibility $\chi(H,T)$ is temperature dependent
($\chi=dm(H)/dH$),
as shown in the Fig.3.

\begin{figure}[b]
\begin{minipage}{1\linewidth}
\includegraphics[width=0.7\linewidth]{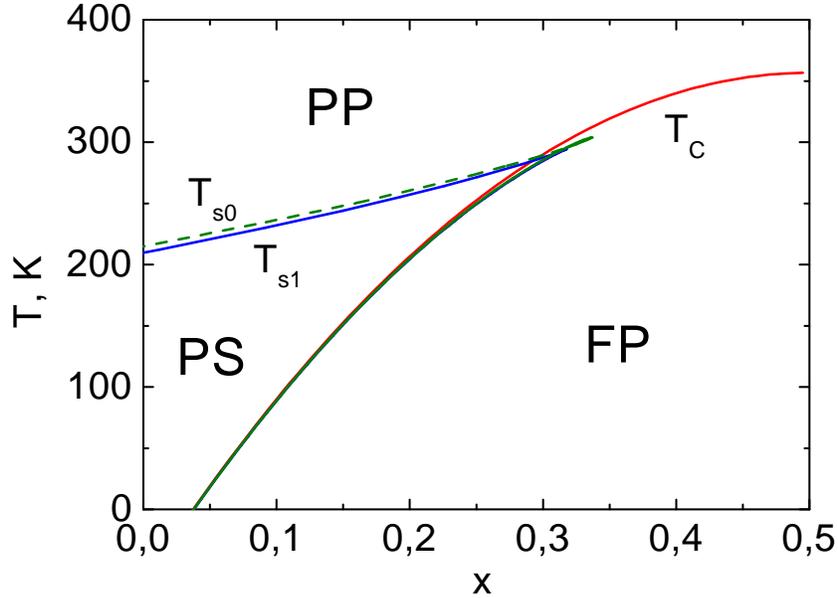}
\caption{Phase diagram of the system with $T_{c0}=-60K$,
$\sigma/\alpha^{'}=1750K$, $Z=2\ 10^6$:
$PP$ is the region of the stable high-temperature phase; $FP$
is the region of the stable ferromagnetic low-temperature phase; $PS$ is the region of the
stable inhomogeneous phase and metastable homogeneous
high-temperature phase.
}
\label{fig:3}
\end{minipage}
\end{figure}

When the external electric field is applied the charged bubbles
shift from equilibrium position. Since the bubbles are not bound to the lattice
they will be accelerated by the field. In real systems bubbles will be pinned to the
lattice by lattice defects and by the Jahn Teller distortions. Therefore the polarization appears
as a result of the shift of the positively charged bubbles with respect
to the negative charged background in the case of hole doping.
Since the bubbles are not strongly bound to the lattice
this shift and the corresponding polarization may be relatively large.
Here we assume that this shift will be limited by Jahn-Teller deformation, which bounds the
bubbles with the lattice.
These arguments may be formulated in terms of relatively simple formula for the effective dielectric constant:
\begin{equation}
\varepsilon_{eff}(T,H)= k \rho_0 (x_1 -\bar{x}) {4\pi \over
3}R_{s}^3
\end{equation}
Here $k$ is the coefficient which determines the local lattice
deformation
which appears due to charge localization and Jahn-Teller effect.
Any charge displacement causes additional local deformation of
the lattice
and therefore protects against large charge displacements. As a
result this
local Jahn-Teller deformation may be estimated as
$d_{JT}(r)\sim     k_{JT} \rho_0 (x(r) -\bar{x})$.
In order to take this effect into account we introduce
the phenomenological coefficient $k$ in Eq.(11) $k\sim k_{JT}^{-1}$.
Note, that in the limit of the strong pinning (for $k_{JT}\to\infty $) 
the charged bubbles becomes strongly bound to the lattice. Coefficient 
$k$ in Eq.(11) becomes small leading to the relatively small 
contribution of bubble displacements to the dielectric permittivity.
The value of the coefficient $k$ very difficult to evaluate theoretically. Therefore
$k$ should be evaluated from the experiments.
But we expect that the pinning is weak.
The value of $k$  should be
relatively large.
The effective dielectric permittivity $\varepsilon_{eff}(T,H)$ (11)
is large because of the large value of $k$
and the large value of polarization.
Importantly $\varepsilon_{eff}(T,H)$
depends on magnetic field, and therefore has magnetoelectric
properties.
The coefficient of the magneto-capacitance effect
$\Delta\varepsilon/\varepsilon(0)=(\varepsilon(H)-\varepsilon(0))/\varepsilon(0)$
may be estimated from Eqs.(7-11) and Fig.2, where the dependence of the effective charge of
the bubbles is plotted as a function of the field. As it follows from Fig. 2 in the
magnetic field in the range 5-7T $\Delta\varepsilon/\varepsilon(0)\simeq 0.5-5$. Note, that magnetic field may cause the
percolation of the charged regions leading to the strong enhancement of the effect.
On the other hand our phenomenological theory is applicable only for the case when different bubbles do not overlap. The discussion of the percolation requires additional theoretical constructions and assumptions and therefore it is out of the scope of our consideration. 

The condition that the inhomogeneous phase has
lower energy then the uniform state ($\Phi_{s}(R_{0},
x_{1})<\Phi_{unif}$) determine the region of the stability of the
inhomogeneous phase. The equation
$\Phi_{s}(R_{s},x_{1})
=\phi_{0}V_0-{H^2/{2\tilde{\alpha}(\bar{x},T)}}$ define the
upper boundary of the existence of the inhomogeneous state.
As a
result this formula and Eqs.(7,8) define
the upper boundary $T_{s1}(\bar{x})$ in the recurrence form as a
function of $x_1$:
\begin{eqnarray}
&&
T_{s1}=T_C+
\frac{\sigma}{\alpha^{'} }x_{1}(1-x_{1})
-
\frac{\sigma Z}
{\alpha^{'} (1-2x_{1})^2}
\\
&& \bar{x}= x_{1} - \frac{2 Z} {3 (1-2x_{1})^3},\ \ \
{\rm where}
\ Z=\frac{16\beta\xi\gamma F}{\pi \sigma^3 }
\end{eqnarray}
These equations  determine the transition temperature $T_{s1}$
to the stable inhomogeneous phase as a function of
external parameters $\bar{x}$. This condition means that
the energy of this inhomogeneous phase is lower then the energy of
homogeneous state. Applying similar procedure we obtain the
equation
which defines the lower boundary of the inhomogeneous phase
$T_{s2}(\bar{x})$. Note that lower boundary $T_{s2}$ will
be always close to the temperature $T_{c\rho}(\bar{x})$, because
there is no any gain in energy, when the
bubble of the high-temperature phase is formed
($\Phi_{\eta}(\eta=0)=0$, and $\Phi_{\eta}(\eta_{0})<0$). It leads
to the essential difference between the formation of the bubble
of the low-temperature phase surrounded by the high-temperature
phase and
the bubble of the high-temperature phase surrounded by the low-
temperature
phase. The first one is energetically favorable and therefore the
region of the existence of these bubbles is considerably large. It
is important to note that Eqs.(11) which determine the phase
diagram does not depend on $d$. This fact allows us to avoid
optimization of the thermodynamic potential with respect to $d$.
Typical phase
diagram of the inhomogeneous state is presented in Fig.4.
Phase transition to the
nonhomogeneous state represents typical first order phase
transition.
Metastable inhomogeneous phase appears at the
temperature $T_{s0}$ which is much higher then the temperature of
phase transition and it is shown in the phase diagram by dashed
line. This line is determined by Eqs.(7,8) and
$A(x_{0})^{2}=4C(x_{0},\bar{x})B(x_{0},\bar{x})$. Phase transition from the phase separated state to a FM phase takes place because the thermodynamic potential of the FM phase becomes lower than the thermodynamic potential of the phase separated state. The FM phase becomes the ground states. Moreover below this temperature the phase separated state does not exist. The minimum of the thermodynamic potential corresponding to inhomogeneous state does not exist any longer. Therefore this phase transition is the phase transition of the first order with the characteristic hysteresis.

Let us compare the calculated phase diagram with the experimental phase diagram of $\rm La_{1-x}Sr_xMnO_3$ \cite{muhin}. The phase transition at $T_s$ corresponds to the transition to inhomogeneous state with the formation of the magnetic long range order at $T_c$ \cite{muhin}. Then with the lowering the temperature the size of bubbles increases and at $T_c$ ($T_p$ in \cite{muhin}) the uniform magnetic state is formed. Note that experimental situation is more complicated. In $\rm La_{1-x}Sr_xMnO_3$ compound at the temperature $T_{O'O''}$, which corresponds to the transition  from weak to strong distorted Jahn-Teller orthorhombic phase,  emerges inhomogeneous fluctuating state. The long-range order between different bubbles is absent in this state. The consideration of this state is out of the scope of our model and therefore it is absent in our phase diagram. Nevertheless we expect that in this state the dielectric permittivity will be large and magnetoelectric effect will be observable as well \cite{mamin}.

In order to make the phase
separation possible it is sufficient to have only small variation
of the charge density per unite cell in comparison with average
charge density
$e(x_{1}-\bar{x})\simeq0.1-0.2e$.
We can estimate the Coulomb contribution to the free
energy $u_{\rho}$ as well as the energy gain due to formation of
the low-temperature phase $u_{\eta}$ (the forth and the second
terms in Eq.(6)) per one unite cell.
For the Coulomb
contribution we obtain:
\begin{equation}
u_{\rho}=\Bigl({3e^{2}(x_{1}-\bar{x})^{2}\over{4\pi \varepsilon
a}}\Bigr) {dR_{0}F\over{a^{2}}}
\end{equation}
Substituting
$\varepsilon=30-40$ (the case of manganites),
$F\simeq 1/18$, $R_{0}\simeq 1-5\ nm (d\sim R_0),$ to Eq.(14) we
obtain that
$u_{\rho}$ is less then $0.02eV$.
Note that because of the screening the Coulomb energy is strongly reduced 
and phase separation becomes possible.
The energy gain due to formation of the
low-temperature phase is $u_{\eta}\simeq 3k_{B}T_{c}$
($u_{\eta}\simeq a^{3}(\tilde{\alpha} (x_{1},T=0))^{2}/\beta\simeq
a^{3}T_{C}\eta_{0}^{2}(x_{1},T=0)/C$,
$C=\eta_{0}^{2}(T=0)/3 k_{B} N_{A}$, where $N_{A}$ is the Avogadro
number). Therefore $u_{\eta}\simeq 0.03-0.1eV >u_{\rho}$ and the
phase separation
becomes favorable.
Therefore the typical size
of the nano-regions, which was estimated from Eq.(7)
as $R_{0}\simeq 1-5\ nm$, represents quite good approximation.

The analysis of the pair distribution function obtained by neutron
scattering shows that the charge density in manganites is
localized on the scale of 3 to 4 interatomic
distances\cite{dagotto,louca}.
The extra
charge in that case is not more then $0.1-0.2e$ per unite cell.
This is consistent with our estimates. This state is
characterized as the state with nano-dimensional charge and phase
separation. Dynamics of
these
charged nano-regions may lead to
high value of the
effective
dielectric constant
$\varepsilon_{eff}(T,H)$ (11) in the low frequency range
\cite{mamin}.

We have shown that the second order phase transition
with
strong dependence of critical temperature
$T_{c\rho}(x)$ (2) on
doping
is unstable with respect to the formation of
the spatially inhomogeneous charged states. Within the
phenomenological Landau theory  we have shown that these states
appears at some temperature $T_{s1}$, which is substantially
higher then the temperature $T_{c\rho}$ (Fig.4). As a result the
phase transition becomes effectively first order phase transition.
Note that the Coulomb interaction determines the charge
distribution, the screening and the characteristic length scale of
the nonhomogeneous states. The spatially inhomogeneous state
becomes possible in the systems with the large dielectric
constants  and with relatively small charge density
variations.
We demonstrated
that  the effective dielectric permittivity $\varepsilon_{eff}(T,H)$
(11)
is large
because of the large value of polarization associated with the shift of bubbles.
And the effective dielectric permittivity $\varepsilon_{eff}(T,H)$ depends on magnetic field.

In conclusion we underline that the localized charged states and
the
phase separation appears even in the case of the second order
phase transition. Properties of these states are described within
the phenomenological theory of the phase transitions. The Coulomb
interaction determines the spatial charge distribution, the
screening and the characteristic length of charge localization.
The inhomogeneous states become possible because of
large dielectric constants and relatively small spatial variation
of the charge density.
These states in the low doped manganites
may lead to the  magnetoelectric behavior.
The giant value of effective dielectric permittivity and the magneto-capacitance effect in the inhomogeneous state
in low-doped systems may become a powerful tool in the investigation
of the inhomogeneous charge segregated states in different materials
including low-doped cuprates
near the threshold the superconducting state.

Enlightening discussions with A.P. Levanyuk, B.Z. Malkin,  and D.
Mihailovic are highly appreciated. We acknowledge financial
support from Slovenian Ministry for Science
and Technology and Ad-Futura (Slovenia).

\end{document}